*Figure 2*

H.Sponholz & D.Molteni

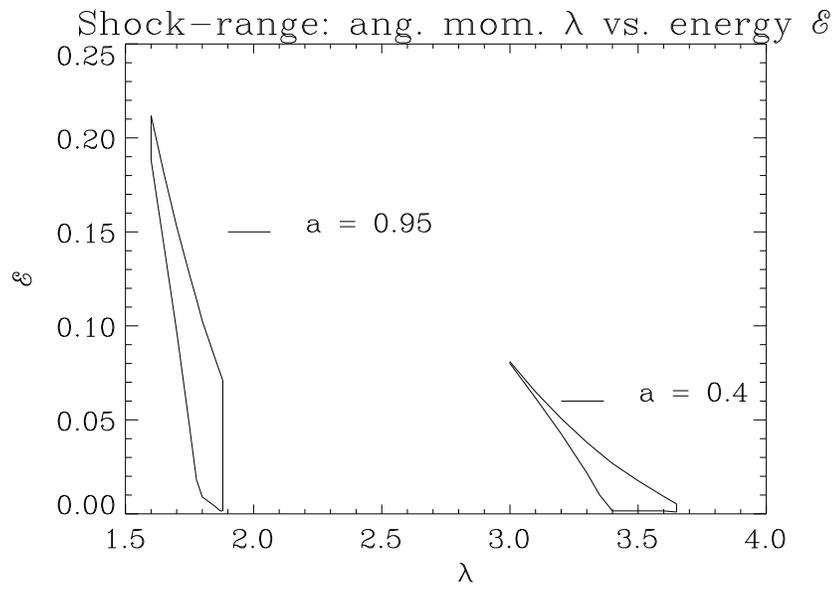

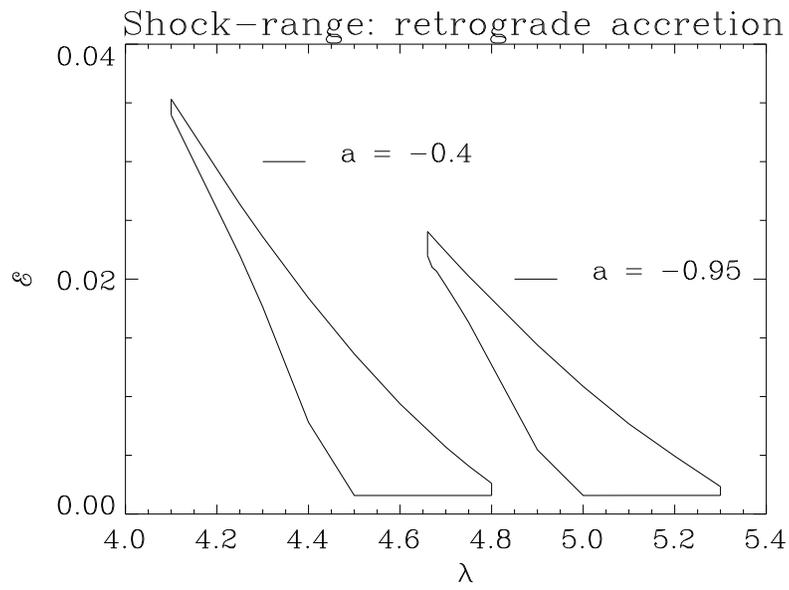

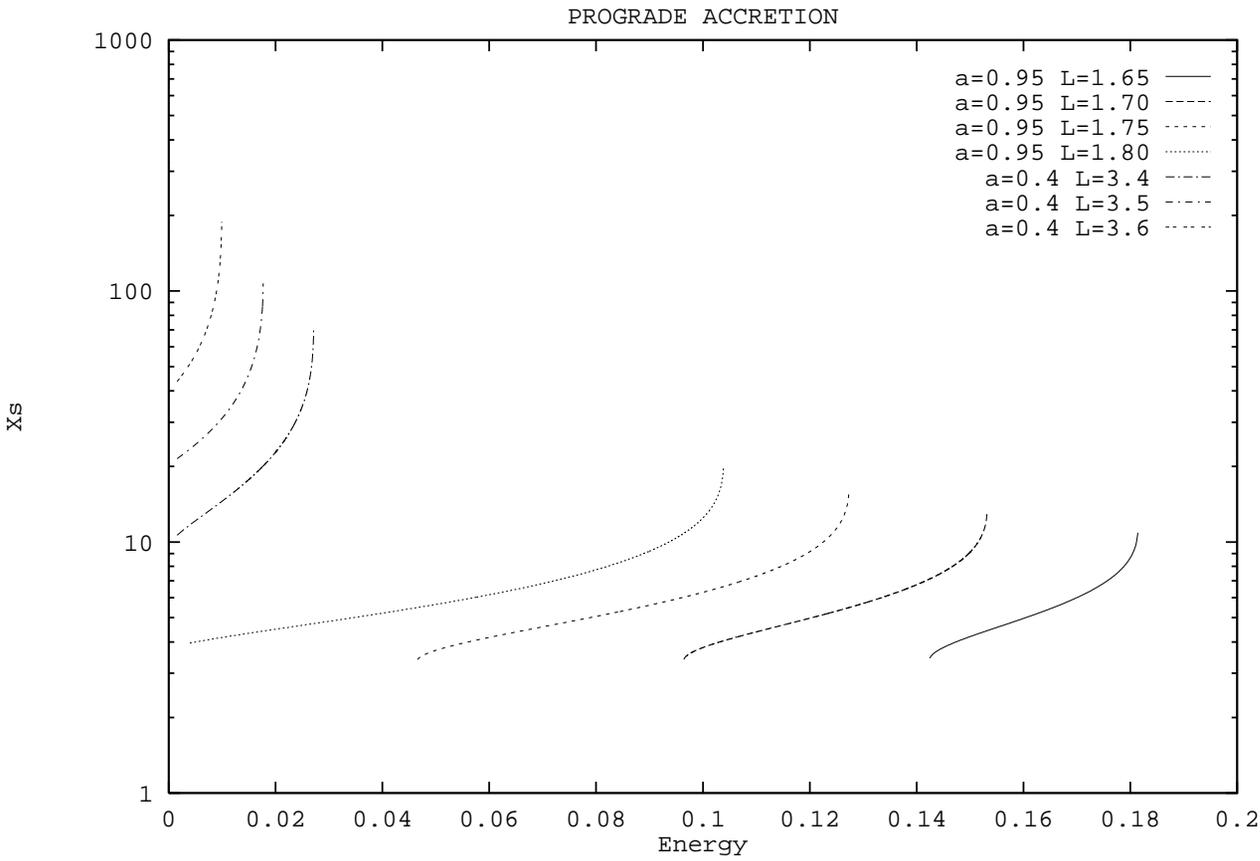

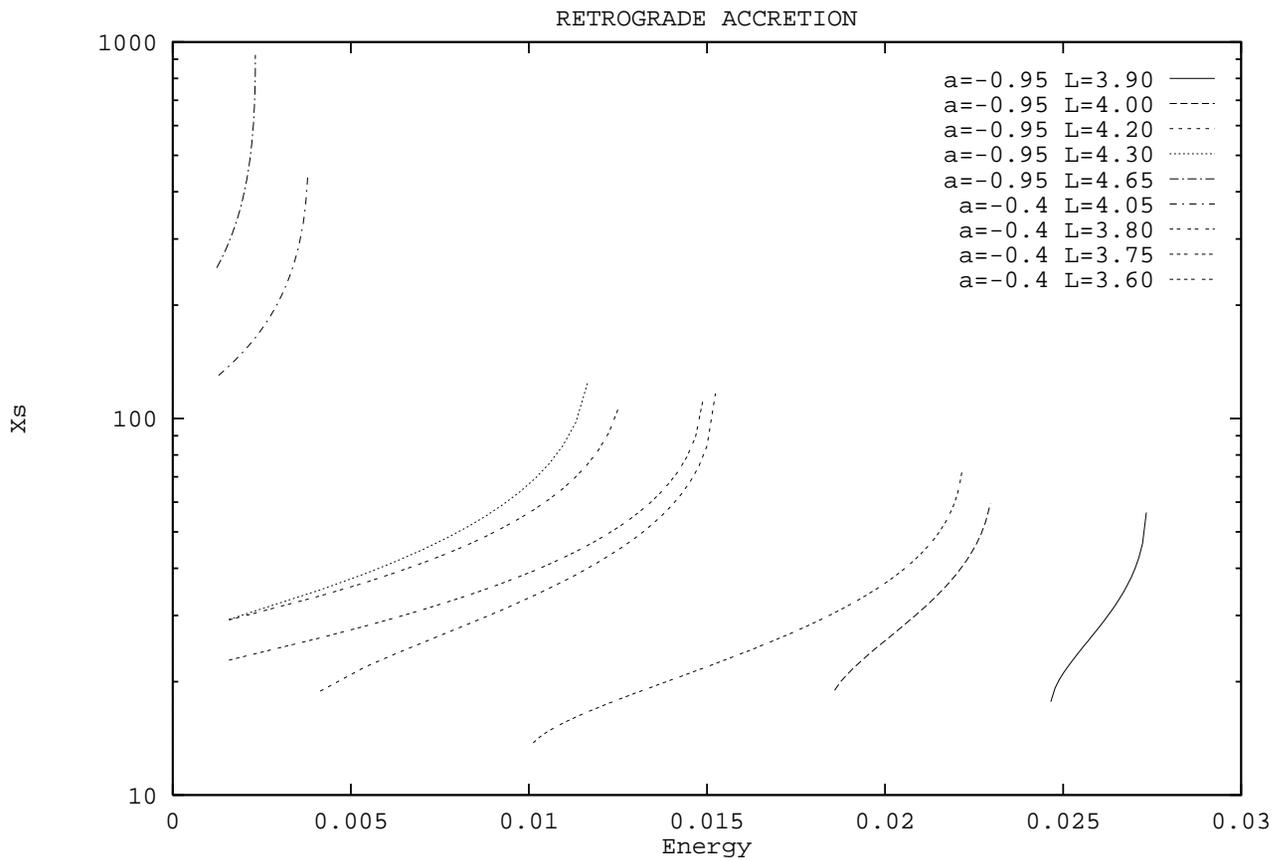

*Figure 5a*

H.Sponholz & D.Molteni

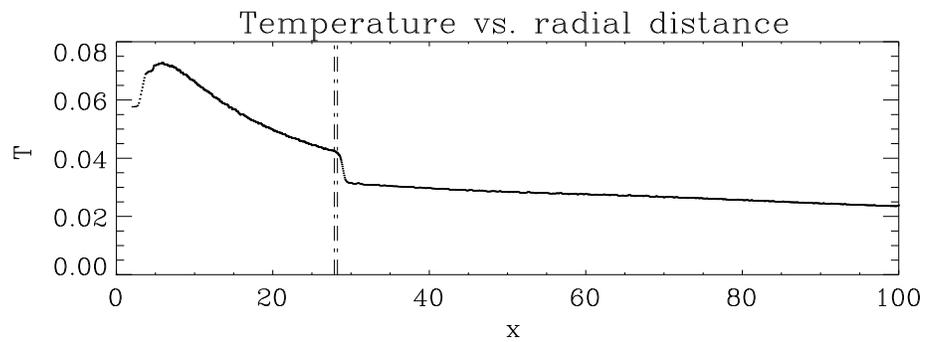

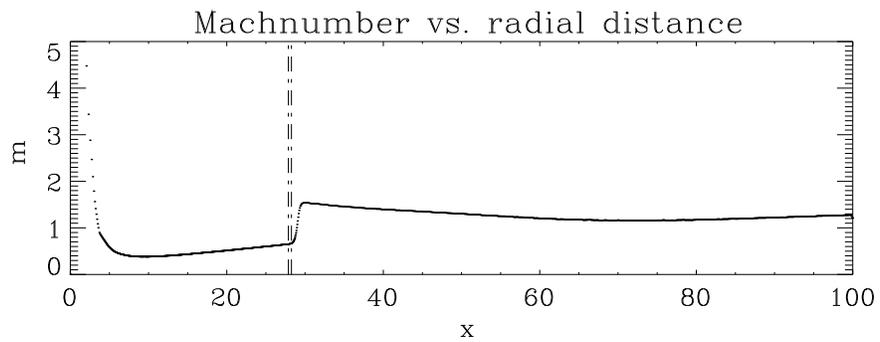

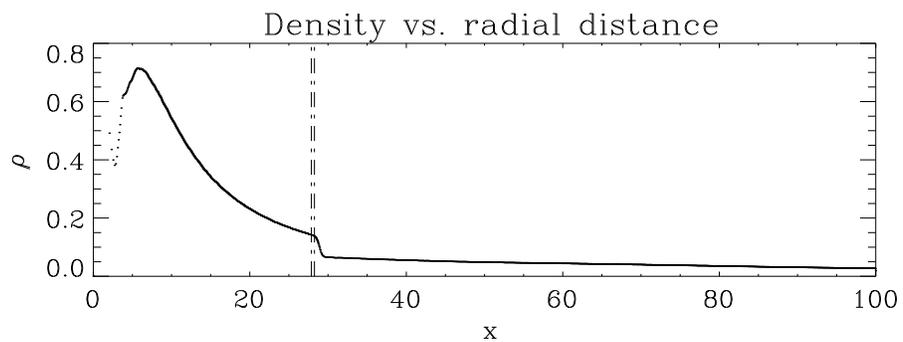

$\lambda = 3.35 \quad a = 0.4 \quad \mathcal{E} = 0.03 \,,\ h = 0.1,\ N = 2008,\ x_s = 28.07$

*Figure 5b*

H.Sponholz & D.Molteni

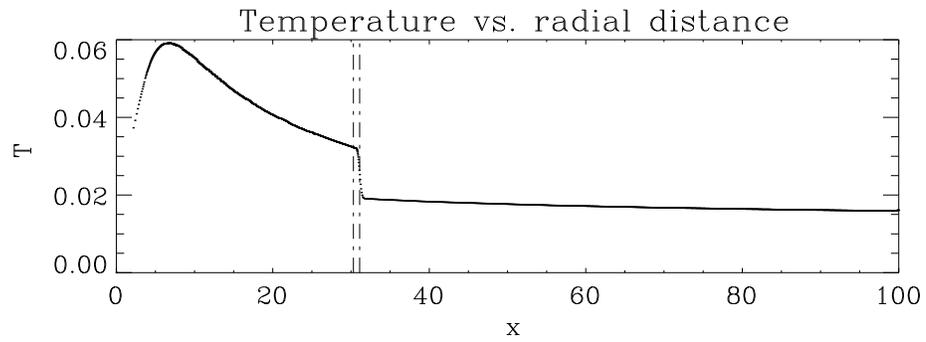

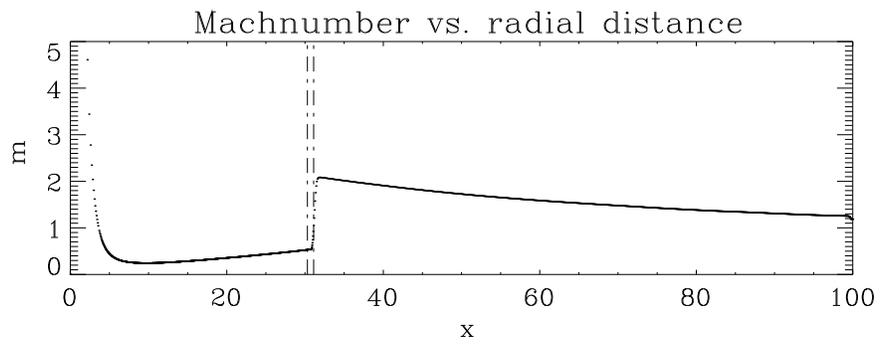

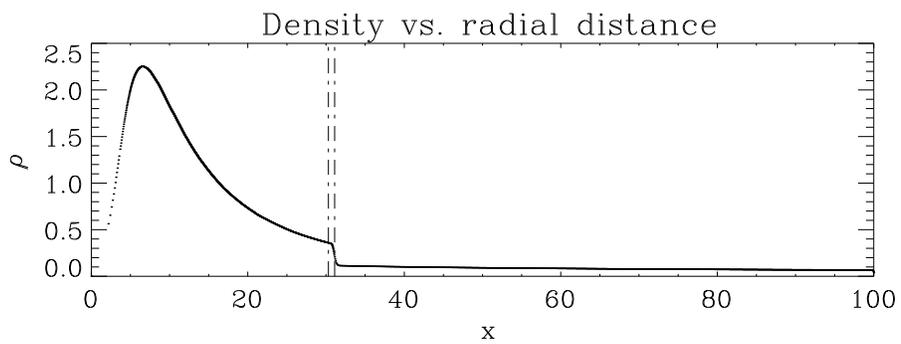

$\lambda$ = 3.45   a = 0.4   $\mathcal{E}$ =0.017 , h= 0.2, N= 2531, $x_s$ =30.70

*Figure 5c*

H.Sponholz & D.Molteni

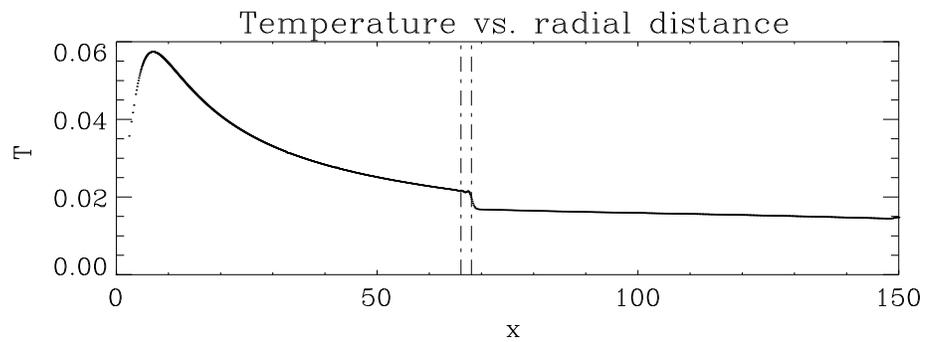

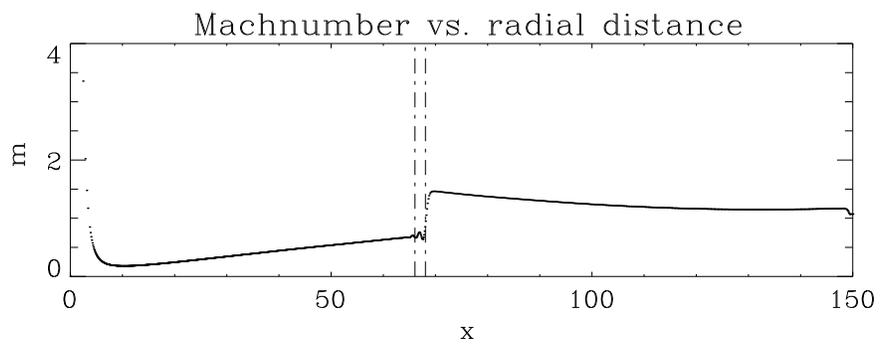

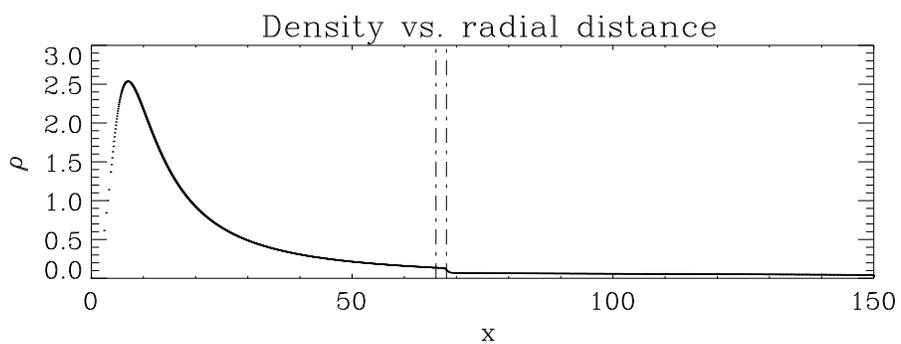

$\lambda = 3.5 \quad a = 0.4 \quad \mathscr{E} = 0.017, \ h = 0.5, \ N = 1843, \ x_s = 67.08$

*Figure 5d*

H.Sponholz & D.Molteni

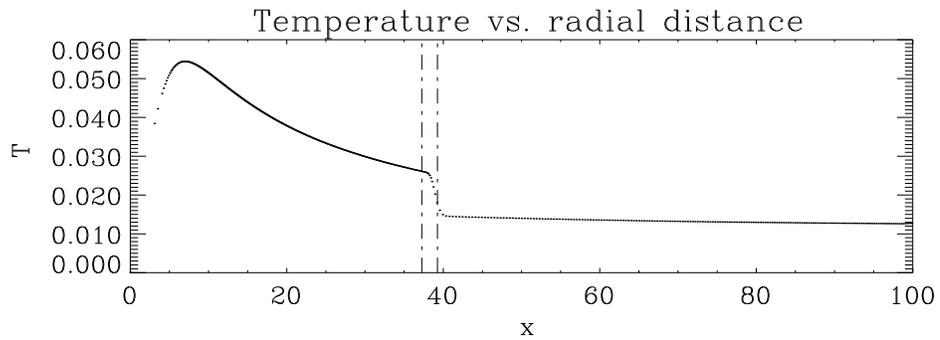

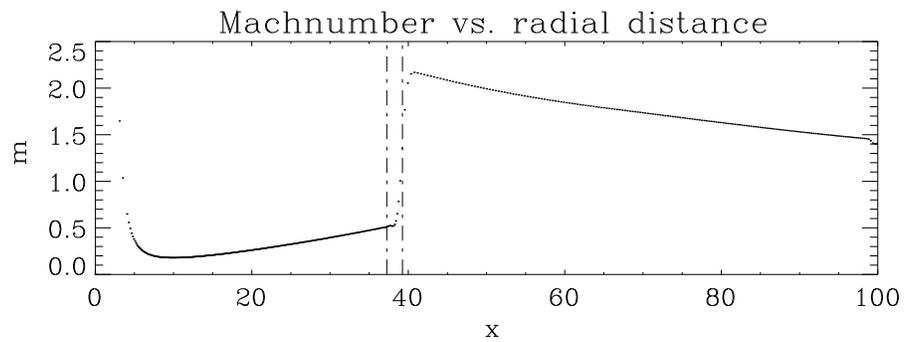

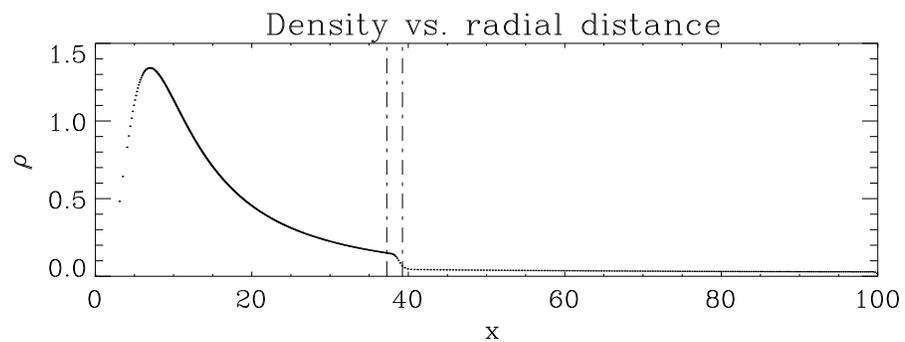

$\lambda = 3.5$  $a = 0.4$  $\mathcal{E} = 0.013$, $h = 0.5$, $N = 576$, $x_s = 38.25$

# Steady state shocks in accretion disks around a Kerr black hole

H. Sponholz[1,2] and D. Molteni[3]
[1] *Institut für Theoretische Astrophysik der Universität Heidelberg, Im Neuenheimer Feld 561, D-69120 Heidelberg, Germany*
[2] *Interdisziplinäres Zentrum für Wissenschaftliches Rechnen, Im Neuenheimer Feld 368, D-69120 Heidelberg, Germany*
[3] *Universita' di Palermo, Istituto di Fisica, Via Archirafi 36, I-90123 Palermo, Italy*

**ABSTRACT**

Results of numerical simulations of shock solutions in a geometrical thin accretion disk around a Kerr black hole (BH) are presented. Using the smoothed particle hydrodynamics (SPH) technique, the influence of the central object is included by means of an effective potential, which is known to mimic the external Kerr geometry very well. We first present the theory of standing shock formation in accretion disks around a Kerr black hole, and show that the results of our numerical simulation agree very well with the theoretical results. Using the pseudo-potential, our analysis gives at least a semi-quantitative impression into what a true general relativistic computation might reveal. The location of the shocks is found to be different for the co-rotating and the counter-rotating flows. Compared to accretion in the vicinity of a non-rotating BH, the position of the shock is shifted inwards for prograde accretion and outwards for retrograde accretion. We find that the shocks in an inviscid flow are very stable. We also remove the ambiguity prevalent regarding the location and stability of shocks in adiabatic flows. Finally we sketch some of the astrophysical consequences of our findings in relation to accretion disks in Active Galactic Nuclei (AGN) and Quasars.

## 1 INTRODUCTION

It is generally believed that the center of active galaxies harbour *rotating* black holes (Rees 1984). The efficiency of conversion of rest mass into radiation, from matter falling onto a rotating black hole, could be as high as forty percent (Shapiro & Teukolsky 1983) and may explain the vast luminosity of active galaxies (Blandford 1990). A Keplerian accretion disk requires a large viscosity for matter to fall in, whereas close to a black hole the accretion time scale could be small compared to the viscous time scale due to very strong gravitational pull resulting from general relativistic effects. Therefore, the specific angular momentum of each fluid element can remain almost constant and the flow may develop steady shock waves for a certain range of parameters (Chakrabarti (1990b) and references therein). Shock waves change the basic temperature structure of the accretion disk and may be responsible for a large number of observed astrophysical phenomena of Active Galactic Nuclei (AGN), such as soft X-ray production and time variability.

In the presented paper, we discuss results of numerical simulations of shock formation in an accretion disk around a Kerr black hole. We use smoothed particle hydrodynamics, which has been well tested for one- and two- dimensional accretion disks (Chakrabarti & Molteni 1993; Molteni, Lanzafame, & Chakrabarti 1994), and the pseudo-Newtonian potential around a Schwarzschild geometry (Paczyński & Wiita 1980).

We use an effective potential for the central object (Chakrabarti & Khanna 1992), which mimics the geometry around a Kerr BH. The Paczyński-Wiita potential mimics the geometry around Schwarzschild Geometry very satisfactorily and a considerable body of the recent astrophysical literature, particularly that dealing with accretion disks, uses this potential with satisfactory results (e.g., Matsumoto et al. (1984), Abramowicz et al. (1988), Kato, Honma, & Matsumoto (1993), Lu (1985), Chakrabarti (1990b), Chakrabarti (1990a)) A similar potential obtained by Chakrabarti and Khanna (1992) is found to reproduce the salient features of hydrodynamic flows around a Kerr geometry quite accurately. Incorporation of these potentials reduces complications in writing numerical codes, without sacrificing the physics, in that a code written in a flat geometry can be trivially modified to study flow behaviour in general relativistic flows. This approach also gives clearer insight into the problem, since the contribution from each term becomes apparent. We thus believe that our approach is well justified and our major conclusions will remain valid even when the full general relativistic equations are solved. At present, the only SPH solutions known are for spherically symmetric Bondi flows (Laguna, Miller, & Zurek 1993). The motivation for doing numerical simulations works stems from the fact that analytical treatments have been carried out only for steady state solutions. Time dependent flows need not choose these solutions. Furthermore, the shock-solutions found in analytical treatments are not necessary stable and may reflect only a transient state. Thus we believe that it is essential to study these solutions using fully numerical time-dependent codes.

For physical reasons shocks solutions exist only in a restricted parameter range in energy and angular momentum (Chakrabarti 1990b). We show that in the case of co-rotating flows, shocks are formed very close to the black hole, whereas for counter-rotating flows the shocks form farther away. Results of our simulation agree very well with the theoretical shock locations which we also present here.

The plan of our paper is the following. In the next section, we discuss the theory of shock formation in a one dimensional accretion flow around a Kerr BH and discuss the regions of parameter space for which shocks are expected to form. The analysis presented is *time independent*. In Section 3 we describe the results of the *time dependent* numerical simulations. Finally, in Section 4, we discuss the relevance of our results in the astrophysical context and present some concluding remarks.

## 2  THEORETICAL SHOCK LOCATIONS IN GEOMETRICALLY THIN ACCRETION FLOWS AROUND A KERR BLACK HOLE

We consider a geometrically thin, adiabatic accretion disk flow in the vicinity of a rotating black hole. This analysis follows the criteria given by Chakrabarti for steady accretion onto Schwarzschild BH (Chakrabarti 1990b; Chakrabarti 1990a). For the accreting matter we use a polytropic equation of state: $P = K\rho^\gamma$; $P$ and $\rho$ denote the isotropic pressure and the matter density, respectively. The adiabatic index ($\gamma$) is assumed to be constant throughout the flow, whereas $K$ is related to the specific entropy ($s$) of the flow. We consider the specific entropy and thus $K$ to be constant everywhere, except at the shock-locations, where entropy-production is allowed. The stationary solution of the problem satisfies the conservation of energy, angular momentum, and mass, supplied by the transonic conditions at the sonic points and the Rankine-Hugoniot conditions at the shock-position. Because the radial flow is pressure driven and caused by the pull of the BH, not by dissipation, the specific angular momentum $\lambda$ remains constant everywhere. Since we are interested in the effects of the rotating black hole, and in particular the frame dragging, we use strictly geometrically thin flow of constant transverse height to avoid complexity. Properties of the shocks are obtained by following the general procedures presented in Chakrabarti (1990b).

### 2.1   The model equations

Although we wish to study hydrodynamical flows around a rotating black hole, we do not consider it necessary to use the exact general relativistic Kerr space-time. Due to the complex nature of the geometry around a Kerr black hole (which introduces additional problems in writing a general relativistic code) we use the pseudo Kerr potential introduced recently (Chakrabarti & Khanna 1992; Khanna & Chakrabarti 1992) to obtain the principle properties of the accretion flow. In a similar manner to which the Paczyński-Wiita potential (Paczyński & Wiita 1980) mimics a non-rotating black hole geometry, the Chakrabarti-Khanna potential mimics a rotating black hole geometry on the equatorial plane quite accurately, and thus enables us to pursue theoretical as well as numerical work using equations in flat geometry. The actual deviation of astrophysically relevant quantities (such as the specific binding energy) predicted by this potential, compared to the exact theory, is very small. Therefore, for all practical purposes, our approach is well justified.

In the following we use as the units of mass, velocity and distance, the mass of the black hole $M_{BH}$, the velocity of light $c$, and the gravitational radius $R_g = GM_{BH}/c^2$, respectively. The radial effective pseudo-potential for the Kerr BH takes the form:

$$g(x) = -\frac{1}{x - x_0} + \frac{x_1 a \lambda}{x^3} + \frac{\omega^2 x^2}{2} + \frac{(1 - 2/x)\lambda^2}{2x^2}$$

with the abbreviations

$$\omega = \frac{2ax}{x^4 + a^2x^2 + 2a^2x} \quad , \quad x_0 = \pm |a|^{\frac{1}{8}} \quad , \quad x_1 = \frac{7}{2} \quad .$$

Here $a$ denotes the Kerr parameter and $x$ is the non-dimensional radial distance from the BH. In this expression a positive spin parameter $a$ represents co-rotating accretion whereas the negative $a$ together with a negative sign for $x_0$ represents counter-rotating accretion. The (constant) angular momentum $\lambda$ of the accretion flow is always taken to be positive. The physical origin of the effective potential and the motivation for the several parameters used, are explained in Chakrabarti and Khanna (1992).

With $\vartheta$ the non-dimensional radial velocity and $c_s$ the sound-speed, the dimensionless energy equation can be written as

$$\mathcal{E} = \frac{\vartheta^2}{2} + \frac{c_s^2}{\gamma - 1} + g(x) \quad . \tag{1}$$

The accretion flow satisfies the equation of mass conservation as given by

$$\dot{m} = \vartheta \rho x h_0 \quad , \tag{2a}$$

where the assumed constant half-thickness is $h_0/4\pi$. We define an entropy sensitive 'accretion rate' $\dot{\mathcal{M}}$ in terms of $\vartheta$, $c_s$ and the polytropic index $n = 1/(\gamma - 1)$:

$$\dot{\mathcal{M}} = \vartheta x c_s^{2n} \quad . \tag{2b}$$

Note that although the flow is separately of constant entropy before and after the shock, entropy increases at the shock and so does $\dot{\mathcal{M}} \sim \dot{m} K^n$.

The physical quantities within the shock region are not resolved on a macroscopic spatial scale much larger than the mean free path of atoms of the gas. At the shock the following (Rankine-Hugoniot) jump conditions for energy, pressure balance and baryon number respectively, are satisfied:

$$\mathcal{E}_+ = \mathcal{E}_- \quad , \tag{3a}$$

$$P_+ + \rho_+ \vartheta_+^2 = P_- + \rho_- \vartheta_-^2 \quad , \tag{3b}$$

$$\dot{m}_+ = \dot{m}_- \quad . \tag{3c}$$

The subscripts '−' and '+' refer to the upstream and the downstream gas properties; equations (3a-c) give the quantities after ('+') the shock in terms of the known values before ('−') the shock.

An accreting flow, initially subsonic at a large distance, has to pass through a critical or sonic point in order to became supersonic and thus enable shock formation. To select the appropriate hydrodynamical solutions the sonic point conditions are derived following the general procedure, described by Chakrabarti (1990a): Differentiating the energy and mass conservation equations (eqns. 1 and 2b) and eliminating $dc_s/dx$ results in the condition,

$$\frac{d\vartheta}{dx}[\vartheta - \frac{c_s^2}{\vartheta}] = \frac{c_s^2}{x} - \frac{dg}{dx} \quad . \tag{4}$$

For a stationary solution with supersonic inner boundary condition, the left-hand side and the right-hand side of equ. (4) vanish simultaneously at the critical points. The vanishing of the left-hand side defines a sonic point,

$$|\vartheta(x_c)| = c_s(x_c) \quad , \tag{5}$$

**Figure 1.** Contours of different constant 'accretion rate' $\dot{\mathcal{M}}$ corresponding to $\mathcal{E} = 0.08, \lambda = 1.8, a = 0.95$. The transonic solution labeled as "SS" represents the physical stationary solution for the $\dot{\mathcal{M}}_- = 1.7891 \times 10^{-4}$, the post-shock solution corresponds to $\dot{\mathcal{M}}_+ = 2.5445 \times 10^{-4}$. The stable shock location is indicated.

whereas vanishing of the right-hand side gives the sound speed at the outer sonic point; the subscript $c$ denotes the quantities at the critical points.

To obtain the full set of solutions which include shock waves, the system of equations (1), (2b), (3a-c), (5) must be solved simultaneously. We do not repeat the general procedure, which has already been discussed in detail in Chakrabarti (1990a). Instead, we present a single example of a solution which includes a standing shock. Fig. 1 presents the contours of constant "accretion rate" $\dot{\mathcal{M}}$ in the Mach number (Mach) versus radial distance (log (x)) plane. The specific energy is chosen to be $\mathcal{E} = 0.08$, the specific angular momentum $\lambda = 1.8$. The Kerr parameter is $a = 0.95$. The solution with $\dot{\mathcal{M}}_- = 1.7891 \times 10^{-4}$ labeled as "SS" represents the physical stationary solution which passes through the the outer sonic point. The post-shock flow with $\dot{\mathcal{M}}_+ = 2.5445 \times 10^{-4}$ passes through the inner sonic point. The jump in $\dot{\mathcal{M}}$ from pre-shock to post-shock value reflects a jump in the entropy function $K$. The stable shock is located at $x = 7.75$ in this case (see also figure 5a). Another shock at $x = 2.28$ is predicted, but it is easily argued that this location is unstable (cp. Chakrabarti & Molteni (1993)).

### 2.2 The parameter space relevant for shock formation

It is well known that shocks can form if a gas meets an obstacle supersonically. In the present case of a black hole accretion, the centrifugal barrier acts as an obstacle to the transonic flow and shocks form in the accretion disk. However, not every set of initial conditions produces a shock. In general, the stronger the potential barrier, the higher the energy required to "push" the matter through the barrier. For the examples presented here, the typical energies required are up to one order of magnitude higher for prograde accretion than for retrograde accretion (fig. 2). It is therefore crucial to have a prior understanding of the regions of pa-rameter space for which stationary shocks may form. Fig. 2 shows some typical regions in $(\mathcal{E}, \lambda)$ space where stationary shocks are both predicted and effectively reproduced by our time-dependent simulations. Results are presented both for co-rotating and counter-rotating flows and for small and large values of the BH rotation parameter $a$.

**Figure 3.** Variation of the locations of the shocks with the energy as a function of various angular momentum (prograde flows).

**Figure 2.** Regions of the parameter space (spanned by $(\mathcal{E}, \lambda)$) where stationary shock solutions are predicted. Upper fig. for co-rotating, and lower fig. for counter-rotating flows. Black hole rotation parameter $a$ is as marked.

**Figure 4.** Variation of the locations of the shocks with the energy as a function of various angular momentum (retrograde flows).

The shocks form closer to the black hole, in the case of prograde accretion flows, than they do in retrograde flows. Fig. 3 shows the variation of the shocks location with energy, for different values of the specific angular momentum. For rapidly rotating black holes, the shocks are usually located less a few tens of the Schwarzschild radii. On the other hand, for slowly rotating black holes ($a = 0.4$) the shocks are located between a few tens and some hundred Schwarzschild radii. Shocks are produced for lower energies and in general for higher angular momentum of the gas flow in the case of counter-rotating black holes; the stationary shocks are located far away from the black hole horizon (Fig. 4). The requirement of higher angular momentum stems from the fact that some of the centrifugal support is 'spent' to fight against the frame-dragging. In general, it is observed that for a fixed $\lambda$ the shock strength, given by the ratio of the pre-shock to the post-shock Mach number, increases as the energy parameter decreases.

## 3  RESULTS OF NUMERICAL SIMULATION OF ACCRETION DISKS

We integrate the classical one dimensional hydrodynamical equations of a compressible inviscid fluid for axial symmetric configurations. Let us denote the cylindrical coordinates by $x$ (radial distance from the $z$-axis of symmetry) and $\phi$. The assumed axial symmetry demands $\frac{d}{d\phi} = 0$. With the Lagrangian derivative

$$\frac{D}{Dt} = \frac{\partial}{\partial t} + v \cdot \nabla$$

we have for the mass conservation

$$\frac{D\varrho}{Dt} = -\frac{\varrho}{x}\frac{\partial(xv_x)}{\partial x} \quad .$$

For the radial and the azimuthal momentum we use

$$\frac{Dv_x}{Dt} = -\frac{v_\phi^2}{x} - \frac{1}{\varrho}\frac{\partial p}{\partial x} - \frac{\partial g(x)}{\partial x} \quad , \quad \frac{Dv_\phi}{Dt} = -\frac{v_\phi v_x}{x},$$

where $g(x)$ describes the effective potential of the Kerr BH as given in sec. 2.1. The energy equation which describes the behavior of the internal energy per unit mass $\epsilon$ is

$$\frac{D\epsilon}{Dt} = -\frac{p}{\varrho}\left[\frac{1}{x}\frac{\partial(xv_x)}{\partial x}\right].$$

The procedure for implementation of the above equations into the SPH formalism is presented in Chakrabarti & Molteni (1993) and is not repeated here.

The general procedure followed in our simulations is presented in Chakrabarti and Molteni (1993); there they showed that the time dependent code reproduces — exactly, within the numerical resolution — the analytical solutions. We use a similar SPH code except that the effective pseudo-Kerr potential (Chakrabarti & Khanna 1992) is used instead of the pseudo-Schwarzschild potential (Paczyński & Wiita 1980). The boundary conditions are as follow: At the inner radius ($r = 1.5 \times GM_{bh}/c^2$) the smoothed particles are simply cut away from the computation; since the flow here is extremely supersonic no feedback disturbances appear on the upstream flow. The smoothed particles are introduced with the azimuthal velocity $v_\phi = \lambda/x_i$ at an injection position ($x_i$) downstream of the outer sonic point where the radial gas speed is already supersonic. The injection velocity $v_x(x_i)$ and the sound speed $c_s(x_i)$ are chosen at this outer boundary according to the prescription for which the theoretical *steady* solution is known from the theory presented in sect. 2.1. A new particle is injected every time an injection sphere $S$ (around the injector position) is free. The flow is smooth for small injection sphere (since many particles overlap). For greater injection areas the flow is more turbulent. We note that for extremely smooth flows the solution obtained in these simulations remains supersonic throughout according to un-shocked solution. When the flow at the outer boundary is more turbulent, the flow switches to the solution which includes a shock. In this case, the flow passes through both the sonic points.

Figures 5, 6 show the shock profiles for different parameters $a$, $\mathcal{E}$, $\lambda$. To within the range of resolution $2h$ ($h$ being the particle size used in the simulation) the shock positions and strengths agree very well with the theoretical results. All the temperature profiles are given in nondimensional units of the thermal energy per unit mass. To estimate the temperature values in Kelvin for this one-dimensional computation an appropriate accretion rate must be chosen and the vertical extent of the flow must be assumed. The accretion rate is expressed in units of the Eddington accretion rate $\dot{M}_E \approx 0.2 M_8 M_\odot year^{-1}$, where $M_8$ denotes the mass of the BH in units of $10^8$ solar masses (Blandford 1990). If we assume the flow at the inflow region to be in vertical force equilibrium and use the resulting vertical extent for the whole radial region, we obtain a temperature presented in figs. 9. If the radiative pressure $P_r$ is much higher than the gas pressure $P_g$ we can assume that the temperature of the 1D disc is given by $T_c^4 = P3c/4\sigma$, where the total pressure $P = P_r + P_g$ is given by the polytropic pressure $P = (\gamma - 1)\varrho U_{th}$. Here the mass density $\varrho = \rho\varrho_0$ is computed from the nondimensional density $\rho$ (which is obtained

**Figure 7.** Theoretical and numerical Mach and temperature profiles. Temperature profiles are scaled in units of the thermal energy per unit mass. The parameters are $a = 0.95, \mathcal{E} = 0.08, \lambda = 1.8$ with the theoretical shock position close to the BH at $x_s = 7.75$. The theoretical shock position within the range of resolution is marked by the two vertical lines. The right parts of both figures show some enlargement of the shock-region. The corresponding temperatures in Kelvin, dependent from the mass accretion ($\dot{M}$) in units of the Eddington accretion rate and the mass of the BH ($M_{BH}$) are displayed in fig. 9, left.

from the computation) and scales with the properties of the accretion flow at the injection point ($x_i$):

$$\varrho_0 = \frac{\dot{M}}{4\pi x_i Z_d v_x(x_i)}$$

with $Z_d$ being the assumed vertical disk equilibrium extent at $x_i$. $U_{th}$ is the thermal energy per unit mass which results from the analytical as well as from the numerical solution, $\gamma = \frac{4}{3}$ denotes the adiabatic index and $\sigma$ denotes the Stefan-Boltzmann constant.

Figures 7, 8 show the excellent agreement between theoretical and numerical Mach number and temperature profiles. These two cases have the parameters (fig. 7) $a = 0.95, \mathcal{E} = 0.08, \lambda = 1.8$ (see also fig. 1), and fig. 8 $a = 0.4, \mathcal{E} = 0.008, \lambda = 3.6$.

Figure 9 shows the corresponding temperature in Kelvin for the choice of various black holes masses and accretion rates. The shock positions agree to within the spatial resolution of the particle size $h$. The temperature and the Mach number agree exceptionally well in the subsonic branch; only in the supersonic branch is there a small difference that it is clearly due to the visible jump at the inflow boundary. The shock starts forming near to the centrifugal barrier and travels backwards with time and eventually become stable practically at the theoretically predicted shock location

**Figure 5.** Prograde flows: Variation of Mach number, temperature profiles (scaled in units of the thermal energy per unit mass), and non-dimensional mass density for different input parameters $a$, $\mathcal{E}$, $\lambda$. The shock positions and strengths agree within the range of resolution ($2h$) with the theoretical predictions. In each figure theoretical shock position with the range of resolution ($x_s \pm 2h$) is marked by the two vertical lines.

Figure 6. Retrograde flows: Variation of Mach number, temperature profiles (scaled in units of the thermal energy per unit mass), and non-dimensional mass density for different input parameters $a$, $\mathcal{E}$, $\lambda$. The shock positions and strengths agree within the range of resolution ($2h$) with the theoretical predictions. In each figure theoretical shock position with the range of resolution ($x_s \pm 2h$) is marked by the two vertical lines.

**Figure 8.** Theoretical and numerical Mach and temperature profiles. Temperature profiles are scaled in units of the thermal energy per unit mass. The parameters are $a = 0.4, \mathcal{E} = 0.008, \lambda = 3.6$ (predicted shock position far away from the BH horizon at $x_s = 80.66$). The theoretical shock position within the range of resolution is marked by the two vertical lines. The right parts of both figures show some enlargement of the shock-region. The corresponding temperatures in Kelvin, dependent from the mass accretion ($\dot{M}$) in units of the Eddington accretion rate and the mass of the BH ($M_{BH}$) are displayed in fig. 9, right.

(c.f. Chakrabarti & Molteni (1993)). In the simulation, we chose $\alpha = 1$, $\beta = 2$ for the artificial viscosity parameters to smooth out the oscillations at both sides of the shock (though a little overshooting is still visible). With a higher artificial viscosity, the shock seems to become diffusive and spreads out on both sides.

In passing, we remark that we have performed several simulations for which theoretical treatments suggest that no shock forms. We do not find any shock formation in the numerical simulations either. Even when large perturbations are introduced, the flow seems to return to its steady supersonic solution after a short period of transients. One of the most significant results of our simulation is that the flow *steady* becomes asymptotically and smoothly passes through *both* the saddle type sonic points simultaneously, without ever shuttling back and forth and causing bi-periodicity as one may naively expect .

## 4 OBSERVATIONAL SIGNIFICANCE AND CONCLUDING REMARKS

In this paper, we have studied the nature of the Rankine-Hugoniot shocks in thin, axisymmetric accretion around a *rotating* black hole both theoretically and through numerical simulation. We find the agreement between these two approaches to be excellent. This provides us with confidence that the code can be trusted not only for thin flows, but for more general accretion flows as well.

Our numerical simulations indicate that shocks may be common in accretion disks, especially if the angular momentum of the disk at a large distance is low compared to the 'Keplerian' value: the angular momentum of the circular orbit around a Kerr BH. This is a possibility if the accretion takes place from winds in star clusters. We have also shown, that out of two theoretically predicted shock locations, only one is stable, as Chakrabarti & Molteni (1993) showed for a nonrotating BH. This removes the ambiguity in shock locations in adiabatic flows.

It is generally believed that the black holes in galactic centers are rotating. One of the ways to be certain that a black hole is rotating is to look for features in a disk which are sensitive to the spin of the black hole. As Chakrabarti (1990a) has pointed out, a stable standing shock could be one such feature. The shocks in co-rotating disks are closer to the hole, which means that the post-shock region of the disk is not very big. In this case, perturbations in the accretion rate should induce faster variabilities in the spectra from the disks, than for accretion by a Schwarzschild BH. Furthermore, the location of the shock itself is very sensitive to the energy and angular momentum of the flow. We suspect that these variabilities may cause the X-ray flickering that is observed in AGNs. On the contrary, the shocks in counter-rotating disks form very far away from the BH, causing a significant part of the disk to be hotter than a Keplerian disk, producing excess UV and possibly soft X-rays. The variability time scale should be longer in this case.

Counter-rotating flows are possible in the case of wind accretions (re-accretion of earlier expelled low-angular momentum winds). A particularly interesting system to observe would be one, where the variability time scale switches (at least once!) from a short time scale to a long time scale (or vice versa). This could represent a case of wind accretion switching from a predominantly co-rotating manner to a predominantly counter-rotating manner (or vice versa). It may be quite possible to observe this, since even a small change in the direction of the wind at a large distance might change the character of the accretion from co-rotating to counter-rotating.

Shocks in accretion disks may be useful in accelerating particles (by, say, first order Fermi processes) and thus producing cosmic rays. It is possible that details of the resulting cosmic ray spectrum strongly depends upon the Kerr parameter $a$. If scatterers (e.g. weak magnetic field) are present close to the shock front in the pre-shock and the post-shock region, particles cross the shock front several times and gain momentum from the scatterers. As the resulting energy spectrum is mainly determined by the difference between the upstream and downstream velocity of the scattering centres, the momentum ($p$) distribution function of the scattered particles is roughly $f(p) \propto p^{-3r/(r-1)}$ where $r$ is the shock compression ratio, the ratio of the upstream to the downstream velocities (Blandford 1990).

We also note that in the case of retrograde accretion, the spatial efficiency of Fermi acceleration of cosmic rays in the standing shock could be higher compared to the same process in a co-rotating BH, or even a Schwarzschild BH, just due to the fact that the shock fronts are more extended in these cases. The detailed dependence of the strength of the shocks on the BH parameters especially for more realistic 2D-simulations, and details of the emitted cosmic ray energy spectrum, will be a topic for future investigation.

**Figure 9.** Temperature profiles in Kelvin for the simulations presented in the figs. different masses ($M_{BH}$) of the BH and different mass accretion rates ($\dot{M}$) related to the Eddington accretion rate $\dot{M}_E$ (cp. sect. 3). For both figs. the parameters are for the graphs from up to down: $M_{BH} = 6M_\odot$: $\dot{M} = 0.1\dot{M}_E$, $\dot{M} = 0.01\dot{M}_E$ and $M_{BH} = 10^4 M_\odot$: $\dot{M} = 0.1\dot{M}_E$, $\dot{M} = 0.01\dot{M}_E$.

### Acknowledgements
We are indebted to Sandip Chakrabarti for helpful discussions and we thank Lewis Ball for a critical reading of the manuscript.
We acknowledge the support by the VIGONI-Program.
H.S. acknowledges support by the Deutsche Forschungsgemeinschaft (Ts 17-3/2).

This paper has been produced using the Blackwell Scientific Publications LaTeX style file.

*Figure 6a*

H.Sponholz & D.Molteni

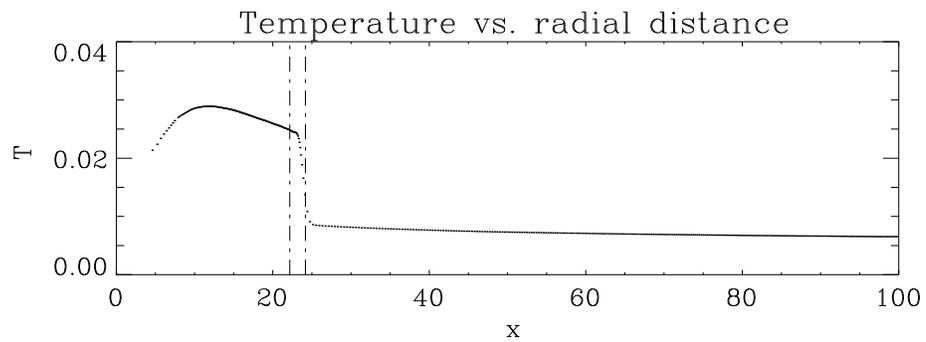

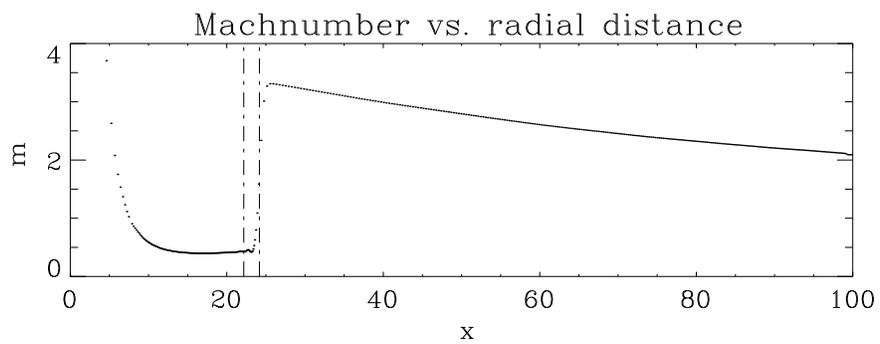

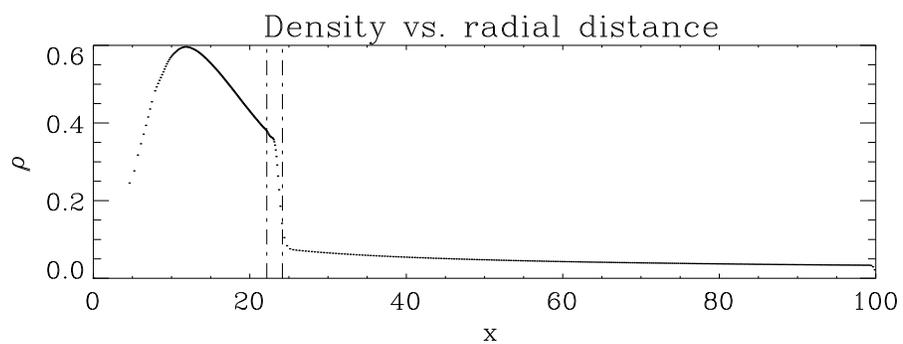

$\lambda = 4.2 \quad a = -0.95 \quad \mathscr{E} = 0.006, \ h = 0.5, \ N = 438, \ x_s = 23.17$

*Figure 6b*

H.Sponholz & D.Molteni

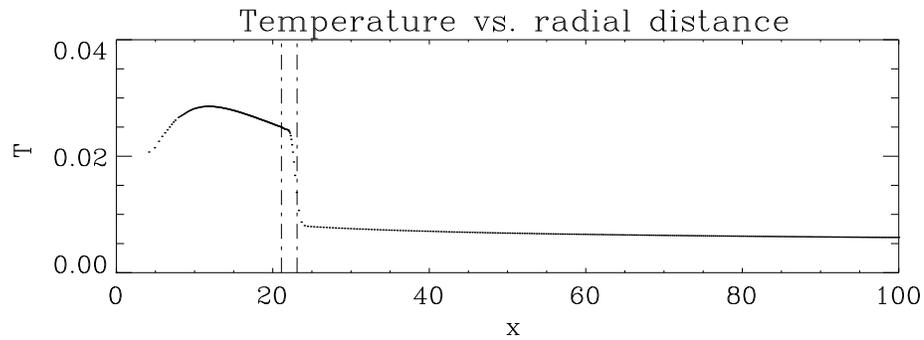

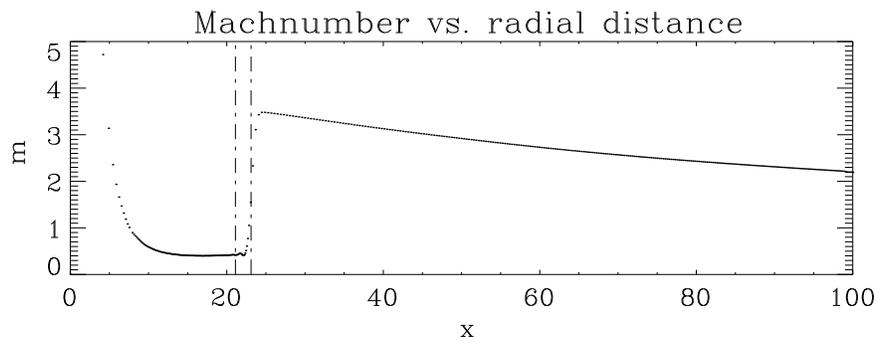

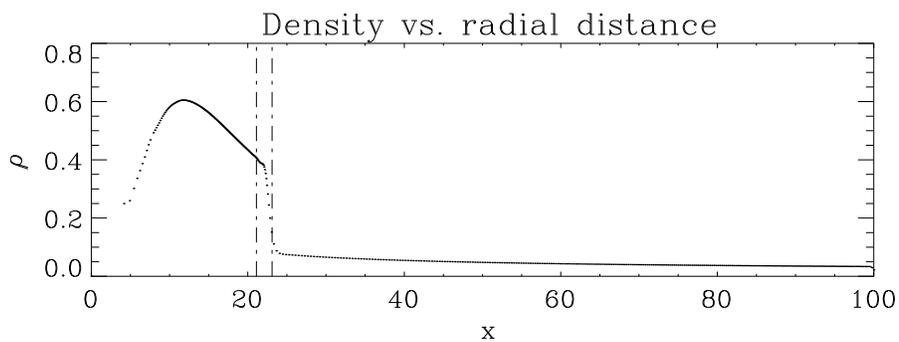

$\lambda = 4.2 \quad a = -0.95 \quad \mathcal{E} = 0.0055 \, , \, h= 0.5, \, N= 433, \, x_s = 22.10$

*Figure 6c*

H.Sponholz & D.Molteni

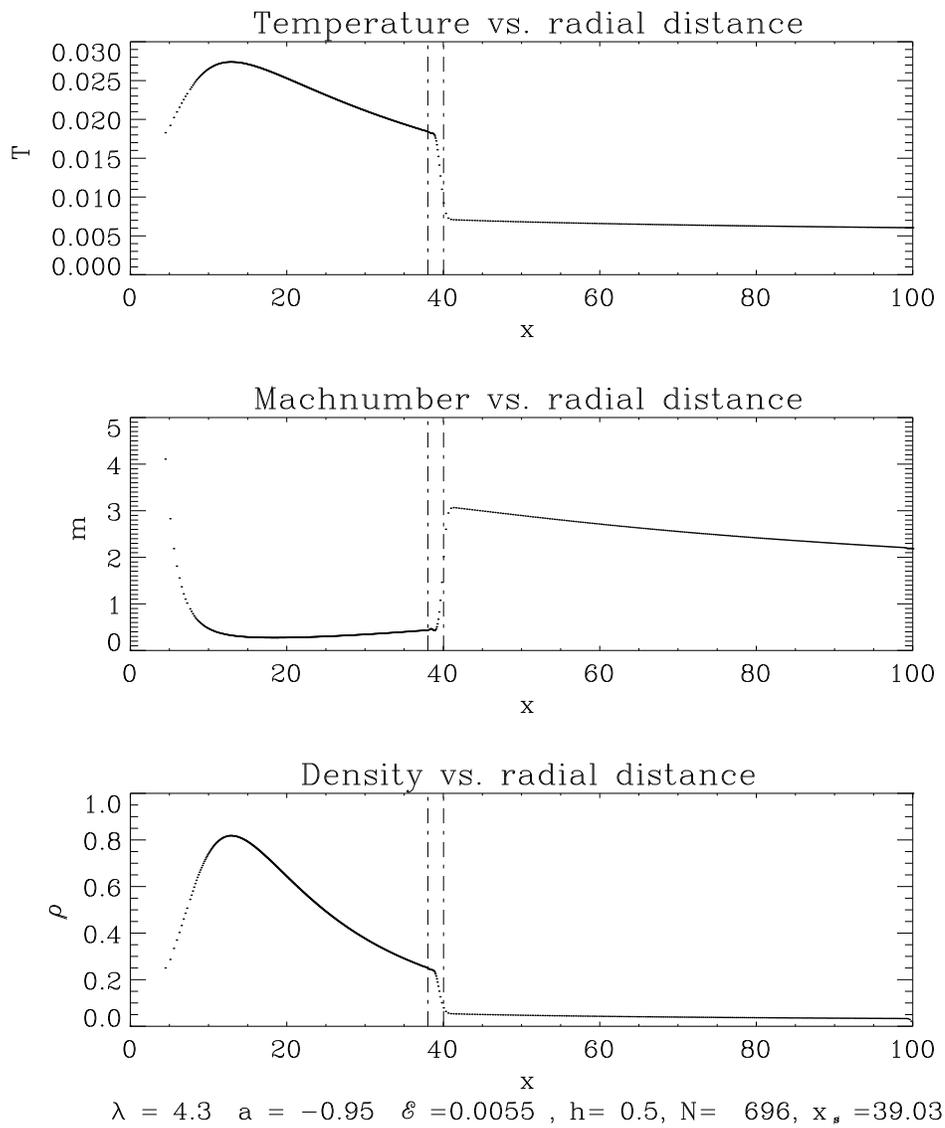

λ = 4.3   a = −0.95   ε =0.0055 , h= 0.5, N=  696, $x_s$ =39.03

*Figure 6d*

H.Sponholz & D.Molteni

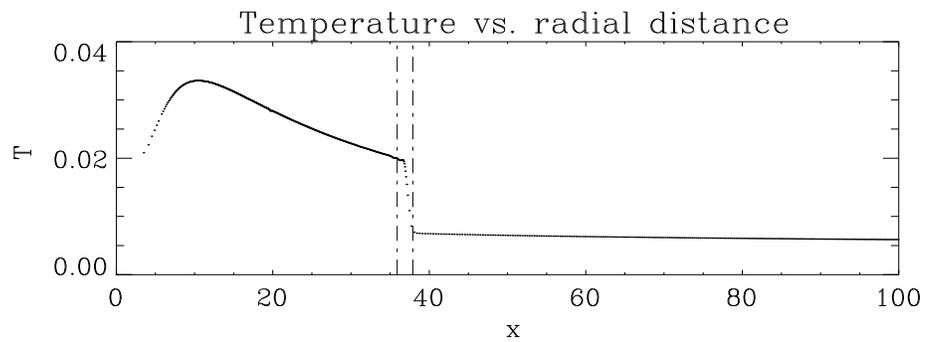

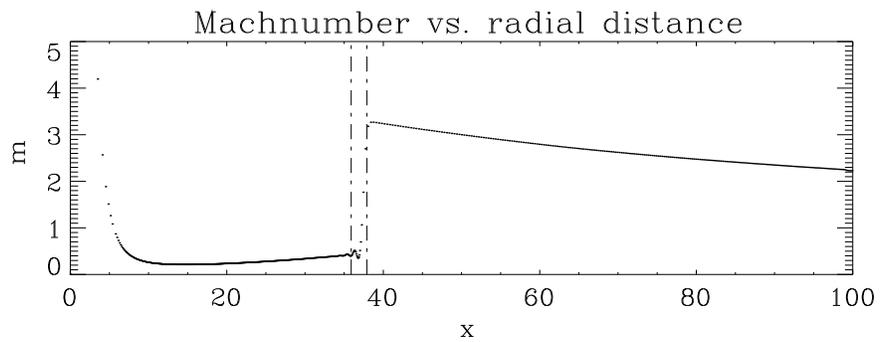

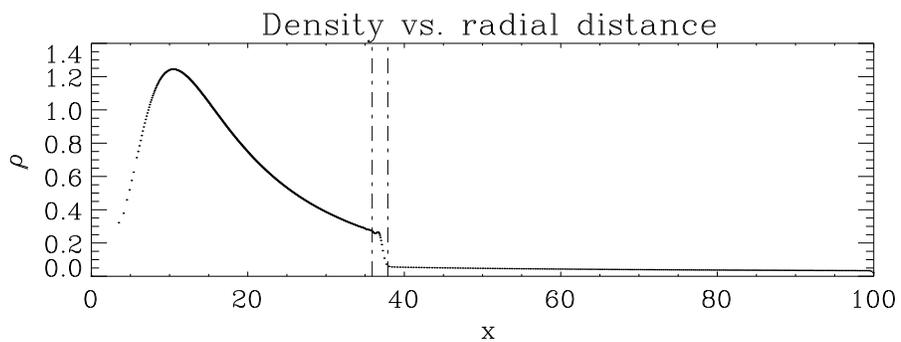

$\lambda$ = 3.8   a = -0.4   $\mathcal{E}$ =0.0055 , h= 0.5, N=  766, $x_s$ =36.90

*Figure 7*

H.Sponholz & D.Molteni

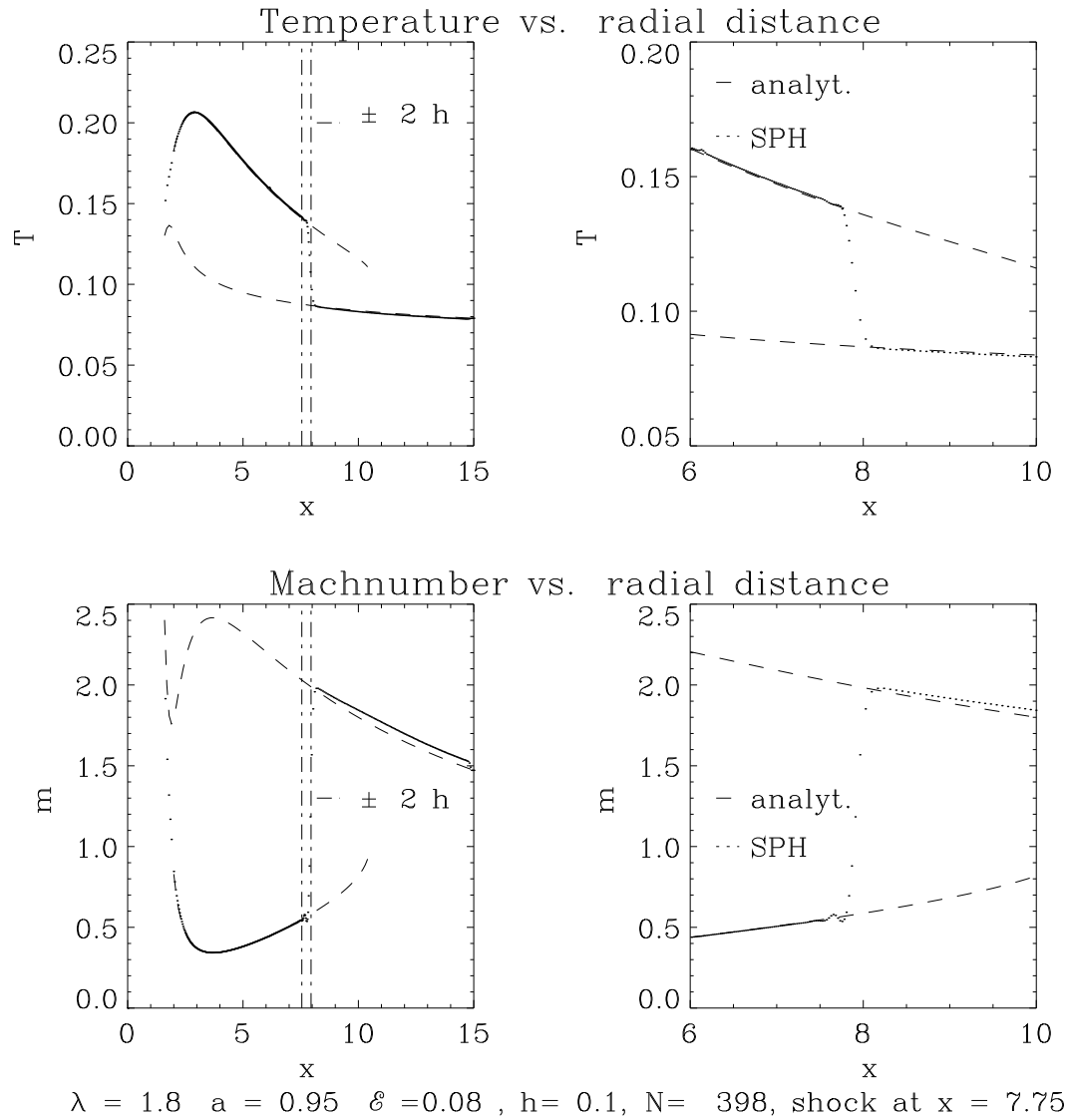

*Figure 8*

H.Sponholz & D.Molteni

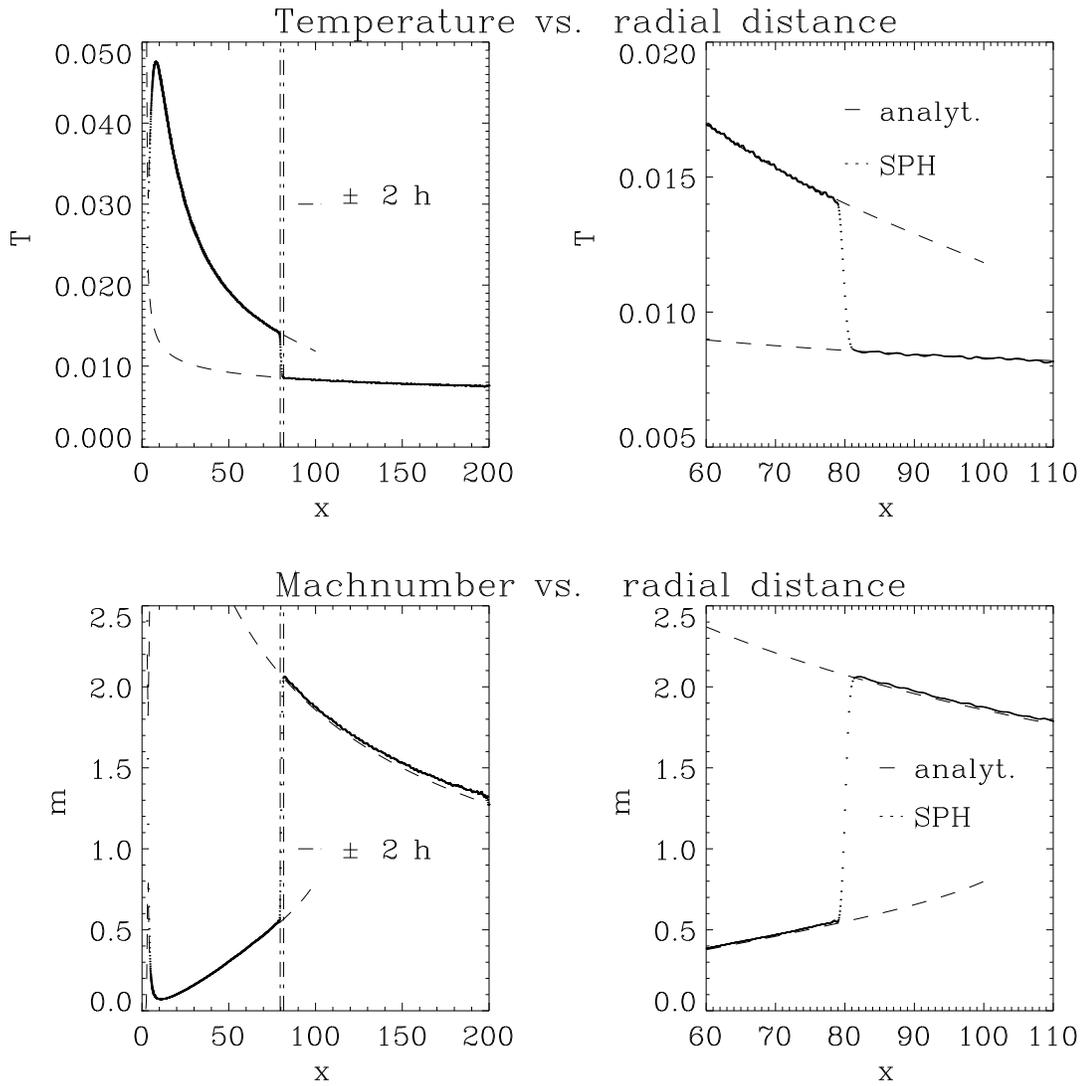

*Figure 9*a

H.Sponholz & D.Molteni

*Figure 9*b

H.Sponholz & D.Molteni

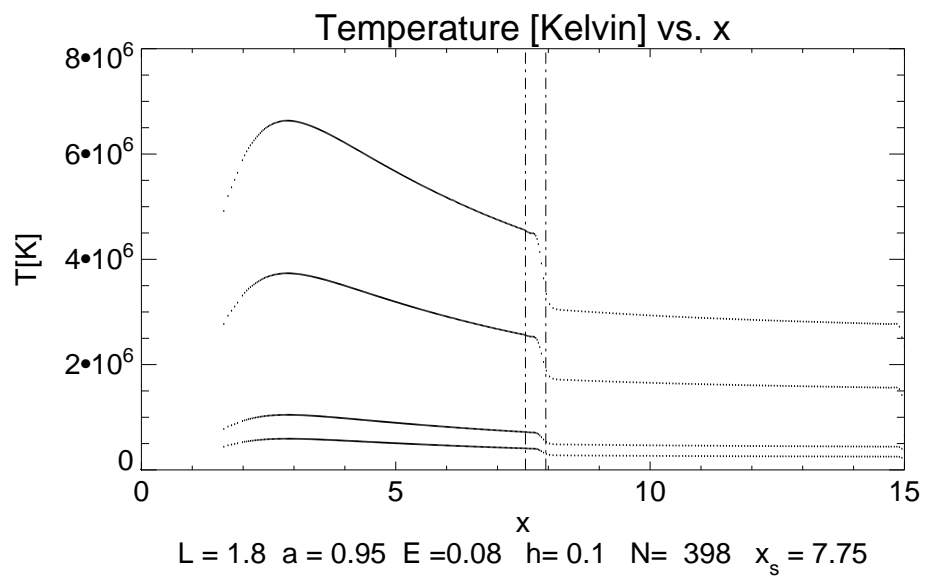

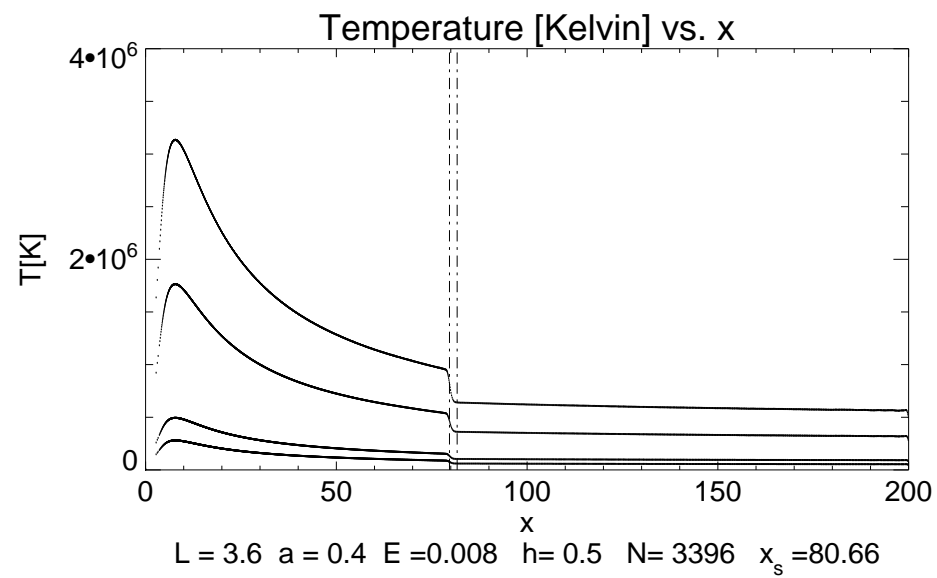